\documentclass[10pt]{article}
\usepackage{latexsym,graphicx}
\newcommand{\be}{\begin{equation}}
\newcommand{\ee}{\end{equation}}
\def\n{\noindent}
\catcode `\@=11
\catcode `\@=12
\begin{document}
\begin{center}
\large{\bf {Kaluza-Klein Type Robertson Walker  Cosmological Model With Dynamical Cosmological 
Term $\Lambda$ }} \\
\vspace{10mm}
\normalsize{ANIRUDH PRADHAN\footnote{Corresponding author}} \\
\vspace{5mm}
\normalsize{Department of Mathematics, Hindu Post-graduate College,
 Zamania-232 331, Ghazipur, India.} \\
\normalsize{E-mail: pradhan@iucaa.ernet.in, acpradhan@yahoo.com}\\
\vspace{5mm}
\normalsize{G. S. KHADEKAR$^2$, VRISHALI PATKI$^3$} \\
\normalsize{Department of Mathematics, R. T. M. Nagpur University, Mahatma Jyotiba Phule
Educational Campus, Amravati Road, Nagpur - 440 033, India.} \\ 
\normalsize{$^2$E-mail: gkhadekar@yahoo.com, gkhadekar@rediffmail.com} \\
\normalsize{$^3$E-mail: patki\_vrushali@rediffmail.com} \\
\vspace{5mm}
\normalsize{SAEED OTAROD$^4$} \\
\normalsize{Department of Physics, Yasouj University, Yasouj, Iran } \\
\normalsize{E-mail: sotarod@mail.yu.ac.ir, sotarod@yahoo.com}
\end{center}
\vspace{10mm}
\begin{abstract}
In this paper we have analyzed the Kaluza-Klein type Robertson Walker (RW) cosmological
models by considering three different forms of variable $\Lambda$: $\Lambda \sim \left
(\frac{\dot{a}}{a}\right)^2$, $\Lambda \sim \left(\frac{\ddot{a}}{a}\right)$ and $\Lambda
\sim \rho$. It is found that, the connecting free parameters of the models with cosmic
matter and vacuum energy density parameters are equivalent, in the context of higher
dimensional space time. The expression for the look back time, luminosity distance and
angular diameter distance are also derived. This work has thus generalized to higher
dimensions the well-known results in four dimensional space time. It is found that there
may be significant difference in principle at least, from the analogous situation in four
dimensional space time.
\end{abstract}
\smallskip
\n PACS: 04.50.+h, 98.80.Es\\
\n Key words : Higher dimensional space time, cosmology, cosmological parameters,
cosmological tests\\
\newpage
\section{Introduction}
Higher dimensional theories of Kaluza-Klein (KK)-type have been considered to
study some aspects of early universe  \cite{ref1} $-$ \cite{ref4}. In such
KK-theory it has been assumed that the extra dimension form a compact manifold of
very small size undetectable at present day energies. Thus in such higher dimensional
theories one would expect that at the grand unification scale the word manifold has
more than one dimension. It is widely believed that a consistent unification of all
fundamental forces in nature would be possible within the space-time with an extra
dimensions beyond those four observed so far. The absences of any signature of extra
dimension in current experiments is usually explained the compactness of extra
dimensions. The idea of dimensional reduction or self compactification fits in
particularly well in cosmology because if we believe in the big bang, our universe
was much smaller at the early stage and the present four dimensional stage could
have been preceded by a higher dimensional one (Chodos and Detweiler \cite{ref1}).
In this work we consider five dimensional Robertson Walker (RW) model as a test case.
In RW type of homogeneous cosmological model, the dimensionality has a marked effect
on the time temperature relation of the universe and our universe appears to cool more
slowly in higher dimensional space time (Chatterjee \cite{ref5}). It has been suggested 
by Marciano \cite{ref6} that  the experimental detection of time variations of 
fundamental constants should be strong evidence for existence of extra dimensions.
Further, developments in super string theory and super gravity
theory have generated great interest in particle physics and
cosmology both to study higher dimensional theory \cite{ref7}.  The string
theory in which the matter fields are confined to a 3-dimensional
``brane-world'' embeded in 1+3+d dimensions, while the gravitational
field can propagate in d extra dimensions, has inspired
researchers and created great interest in higher- dimensional
gravity theories. The brane world model presented by Randal and
Sundrum \cite{ref8}, has attracted much attention as an alternative to the
standard four dimensional cosmology because standard four
dimensional gravity is recovered on the brane in the low energy
limit \cite{ref9}. Due to the posibility of the large extra-dimensions in
the brane-world models, the cosmological models of the early
universe must be studied with careful consideration of the effect
of the bulk geometry \cite{ref10}. Recently, Sasaki \cite{ref11} has presented
review on recent progress in brane -world cosmology. Very
recently, Alam and Sahni \cite{ref12} have compared the predictions of
brane-world models to two recently released Supernova data sets:
the Gold data \cite{ref13} and the data from the Supernova Legacy Survey
(SNLS)\cite{ref14}. They have shown that brane-world models satisfy both
sets of S Ne data. In recent years several authors 
(see Refs. \cite{ref15} $-$ \cite{ref17}
and references therein) have studied multidimensional
inhomogeneous cosmological models in diferent context.

In recent year, models with a relic cosmological constant $\Lambda$
have received considerable attention among researchers for various reasons
(see Refs. \cite{ref18} $-$ \cite{ref20} and references therein). We should
realize that the existence of a nonzero cosmological constant in Einstein's
equations is a feature of deep and profound consequence. The recent
observations indicate that $\Lambda \sim 10^{-55}cm^{-2}$ while particle physics
prediction for $\Lambda$ is greater than this value by a factor of order $10^{120}$.
This discrepancy is known as cosmological constant problem. Some of the recent
discussions on the cosmological constant ``problem'' and
consequence on cosmology with a time-varying cosmological constant are investigated by
Ratra and Peebles \cite{ref21}, Dolgov \cite{ref22} $-$ \cite{ref24},
Sahni and Starobinsky \cite{ref25}, Padmanabhan \cite{ref26} and Peebles \cite{ref27}.
For earlier reviews on this topic, the reader is referred to Zeldovich \cite{ref28},
Weinberg \cite{ref29} and Carroll, Press and Turner \cite{ref30}.

Recent observations of type Ia supernovae (SNe Ia) at redshift $z < 1$  provide startling
and puzzling evidence that the expansion of the universe at the present time appears to be
{\it accelerating}, behaviour attributed to ``dark energy'' with negative pressure.
These observations (Perlmutter {\it et al.} \cite{ref31}; Riess {\it et al.} \cite{ref32};
Garnavich {\it et al.} \cite{ref33}; Schmidt {\it et al.} \cite{ref34}) strongly favour
a significant and positive value of $\Lambda$. The main conclusion of these observations is
that the expansion of the universe is accelerating.

A number of authors have argued in favour of the dependence $\Lambda \sim t^{-2}$ first
expressed by Bertolami \cite{ref35} and later on by several authors \cite{ref36} $-$ \cite{ref41}
in different context. Recently, motivated by dimensional grounds in keeping with quantum cosmology,
Chen and Wu \cite{ref42}, Abdel-Rahaman \cite{ref43} considered a $\Lambda$ varying as $a^{-2}$.
Carvalho {\it et al.} \cite{ref44}, Waga \cite{ref45}, Silveira and Waga \cite{ref46}, Vishwakarma
\cite{ref47} have also considered/modified the same kind of variation. Such a dependence alleviates
some problems in reconciling observational data with the inflationary universe scenario. Al-Rawaf
and Taha and Al-Rawaf \cite{ref48} and Overdin and Cooperstock \cite{ref49} proposed a cosmological
model with a cosmological constant of the form $\Lambda = \beta \frac{\ddot{a}}{a}$, where $\beta$
is a constant. Following the same decay law recently Arbab \cite{ref50} have investigated cosmic
acceleration with positive cosmological constant and also analyze the implication of a model
built-in cosmological constant for four-dimensional space time. The cosmological consequences
of this decay law are very attractive. This law provides reasonable solutions to the cosmological
puzzles presently known. One of the motivations for introducing $\Lambda$ term is to reconcile
the age parameter and the density parameter of the universe with recent observational data.

Vishwakarma \cite{ref51} has studied the magnitude-redshift
relation for the type Ia supernovae data and the angular
size-redshift relation for the updated compact radio sources data
\cite{ref52} by considering four variable
$\Lambda$-models:$\Lambda \sim a^{-2}$, $\Lambda \sim H^{-2}$,
$\Lambda \sim \rho$ and $\Lambda \sim t^{-2}$. Recently Ray and
Mukhopadhyay \cite{ref53} have solved Einstein's equations for
specific dynamical models of the cosmological term $\Lambda$ in
the form: $\Lambda \sim \left(\frac{\dot{a}}{a}\right)^2$,
$\Lambda \sim \left(\frac{\ddot{a}} {a}\right)$ and $\Lambda \sim
\rho$ and shown that the models are equivalent in the framework of
flat RW space time.

In this context, the aim of the present work is based on recent
available observational information. In this paper the implication
of cosmological model with cosmological term of three different
forms: $\Lambda = \alpha \left(\frac{\dot{a}}{a}\right)^2$,
$\Lambda = \beta \left(\frac{\ddot{a}} {a}\right)$ and $\Lambda =
8\pi\gamma \rho$, where $\alpha$, $\beta$ and $\gamma$ are free
parameters, are analyzed within the framework of higher
dimensional space time. The paper is organized as follows. The
Einstein's field equations and their general solutions for
different $\Lambda$-dependent models are presented in Sections $2$
and $3$ respectively. The equivalence of different forms of
dynamical cosmological term $\Lambda$ are discussed in Section
$4$. The cosmological tests pertaining neoclassical test,
luminosity distance, angular diameter distance and look back time
for the model are discussed in Sections $5-8$. As a concluding
part some discussions are made in Section $9$.

\section{The Metric and Field  Equations}
We consider the Kaluza-Klein type Robertson Walker (RW) space time
\begin{equation}
\label{eq1}
ds^{2} = dt^2 - a^{2}(t)\left[\frac{dr^2}{1-k r^2}+
r^2 (d\theta^{2} + \sin^{2}\theta d\phi^{2}) + (1 - kr^{2})d\psi^{2}\right],
\end{equation}
where $a(t)$ is the scale factor, $k=0,\; \pm 1$ is the curvature
parameter. \\
The usual energy-momentum tensor is modified by addition of a term
\begin{equation}
\label{eq2}
T^{vac}_{ij} = - \Lambda(t) g_{ij},
\end{equation}
where $\Lambda(t)$ is the cosmological term and $g_{ij}$ is the
metric tensor. \\

Einstein's field equations (in gravitational units $c = 1, \; \; G = 1$) read as
\begin{equation}
\label{eq3}
 R_{ij} - \frac{1}{2} R g_{ij} = - 8 \pi
T_{ij}-\Lambda(t) g_{ij}.
\end{equation}
The energy-momentum tensor $T_{ij}$ in the presence of a perfect fluid has the form
\begin{equation}
\label{eq4}
T_{ij} = (p + \rho)u_{i}u_{j} - p g_{ij},
\end{equation}
where $p$ and $\rho$ are, respectively, the energy and
pressure of the cosmic fluid, and $u_{i}$ is the fluid
five-velocity such that $u^{i}u_{i} = 1$. \\
The Einstein filed Eqs. (\ref{eq3}) and (\ref{eq4}) for the metric (\ref{eq1})
take the form
\begin{equation}
\label{eq5}
6\left(\frac{\dot{a}^{2}} {a^{2}} + \frac{k}{a^2}\right)=8 \pi\rho + \Lambda(t),
\end{equation}
\begin{equation}
\label{eq6}
\frac{3\ddot{a}}{a} + 3\left(\frac{\dot{a}^{2}}{a^{2}}+\frac{k}{a^2}\right) = - 8 \pi p + \Lambda(t).
\end{equation}
An over dot indicates a derivative with respect to time $t$. The energy conservation law can be
written as
\begin{equation}
\label{eq7}
\dot{\rho} + 4(\rho + p)H = -\frac{\dot{\Lambda}}{8\pi}.
\end{equation}
where $H = \frac{\dot{a}}{a}$, is the Hubble parameter.

For complete determinacy of the system, we consider the barotropic equation of state

\begin{equation}
\label{eq8}
p = \omega \rho,
\end{equation}
where equation of state parameter $\omega$ can take the constant values $0$, $1/3$, $-1$ and $+1$
respectively for the dust, radiation, vacuum fluid and stiff fluid.

By using Eqs. (\ref{eq8}) and (\ref{eq5}),  Eq. (\ref{eq6})
reduces to
\begin{equation}
\label{eq9}
\frac{\ddot{a}}{a} + \frac{4\pi\rho}{3}(2\omega + 1) = \frac{\Lambda}{6}.
\end{equation}
By using Eqs. (\ref{eq5}), (\ref{eq6}) and (\ref{eq8})  for eliminating $\rho$, we obtain
\begin{equation}
\label{eq10}
\frac{3\ddot{a}}{a} + 3(1 + 2\omega)\left(\frac{\dot{a}^{2}}{a^{2}}+\frac{k}{a^2}\right) =
(1 + \omega)\Lambda(t).
\end{equation}
This is the dynamical equation related to the cosmic scale factor $a$ for a known value of the
dynamical cosmological term $\Lambda$. It can be readily be observed from the above equations
(9) and (10) that $\Lambda$ is dependent on the factor $\frac{\ddot{a}}{a}$, $\rho$,
$\left(\frac{\dot{a}}{a}\right)^{2}$ and $a^{-2}$ in specific way. However, inflationary theory
of universe predicts and CMB detectors such as BOOMERANG (de Bernardis {\it et al.} \cite{ref54};
Netterfield {\it et al.} \cite{ref55}), MAXIMA (Hanany {\it et al.} \cite{ref56}; Lee {\it et al.}
\cite{ref57}; Balbi {\it et al.} \cite{ref58}), DASI (Halverson {\it et al.} \cite{ref59}), CBI
(Sievers {\it et al.} \cite{ref60}) and WMAP (Bennett {\it et al.} \cite{ref61}; Spergel {\it et al.}
\cite{ref56}) confirm that the universe is spatially flat. Therefore, $\Lambda \sim a^{-2}$ case
does not sustain for $k = 0$ and therefore omitted here. Hence we shall consider the cosmological
term related to the cases  $\left(\frac{\dot{a}}{a}\right)^{2}$, $\frac{\ddot{a}}{a}$ and
$\rho$ and obtained the solutions in the following sections.
\section{Three Different Forms of $\Lambda$}
\subsection{Model for $\Lambda \sim \left(\frac{\dot{a}}{a}\right)^{2}$}

If we use $\Lambda = \alpha\left(\frac{\dot{a}}{a}\right)^{2}$ = $\alpha H^{2}$, where $\alpha$
is a constant, then for flat universe $k = 0$ the Eq. (\ref{eq10}) reduces to
\begin{equation}
\label{eq11}
a\ddot{a} + \frac{1}{3}\left[\omega(6 - \alpha) + (3 - \alpha)\right]\dot{a}^{2} = 0.
\end{equation}
Integrating Eq. (\ref{eq11}), we get general solution as
\begin{equation}
\label{eq12}
a(t) = \left[\frac{1}{3}(6 - \alpha)(1 + \omega)k_{1}t\right]^{\frac{3}{(6 - \alpha)(1 + \omega)}},
\end{equation}
\begin{equation}
\label{eq13}
\rho(t) = \frac{9}{8\pi(6 - \alpha)(1 + \omega)^{2}}\frac{1}{t^{2}},
\end{equation}
\begin{equation}
\label{eq14}
\Lambda(t) = \frac{9\alpha}{(6 - \alpha)^{2}(1 + \omega)^{2}}\frac{1}{t^{2}},
\end{equation}
where $k_{1}$ is an integrating constant. \\
From the above set of equations (\ref{eq12}) - (\ref{eq14}) it is to be noted that $\alpha \ne 6$
for physical validity. From Eq. (\ref{eq13}), it is observed that $\rho > 0$ implies $0 < \alpha
< 6$. The case $\alpha \geq 6$ is either unphysical or not compatible with time dependent $\Lambda$.
As $\Lambda \ne 0$, then $\dot{a} \ne 0$, which suggests that when $\Lambda \sim \left(\frac{\dot{a}}
{a}\right)^{2}$, then as long as $\Lambda \ne 0$, the expansion of the universe stops.
\subsection{Model for $\Lambda \sim \frac{\ddot{a}}{a}$}

If we consider  $\Lambda = \beta\frac{\ddot{a}}{a}$, where $\beta$
is a constant, then for flat universe $k = 0$ the Eq. (\ref{eq10}) reduces to
\begin{equation}
\label{eq15}
a\ddot{a} + \frac{3(1 + 2\omega)}{(3 - \omega \beta - \beta)}{\dot{a}}^{2} = 0.
\end{equation}
Equation (\ref{eq15}) gives the general solution as
\begin{equation}
\label{eq16}
a(t) = \left[\frac{(1 + \omega)(6 - \beta)}{(3 - \omega \beta - \beta)}k_{2}t\right]^{\frac{
(3 - \omega \beta - \beta)}{(1 + \omega)(6 - \beta)}},
\end{equation}
\begin{equation}
\label{eq17}
\Lambda(t) = \left[\frac{3\beta(1 + 2\omega)(\beta + \omega\beta - 3)}{(1 + \omega)^{2}
(\beta - 6)^{2}}\right]\frac{1}{t^{2}},
\end{equation}
\begin{equation}
\label{eq18}
\rho(t) = \frac{3(\beta + \omega \beta - 3)}{8\pi(\beta - 6)(1 + \omega)^{2}}\frac{1}{t^{2}},
\end{equation}
where $k_{2}$ is an integrating constant.

It is remarkable to mention here from Eqs. (\ref{eq16}) - (\ref{eq18}) that for physical valid
solutions $\beta \ne 0$ and also $\beta \ne 6$. It is also observed from above equations that
for $\beta > 3$, $\Lambda >0$ and for $\beta > 6$, $\rho > 0$ for dust model. Thus, either
$\beta < 0$ or $\beta > 6$ for our present model.
\subsection{Model for $\Lambda \sim \rho$}

If we set $\Lambda = 8\pi \gamma \rho$, where $\gamma$ is a free parameter, then using
Eq. (\ref{eq5}), Eq. (\ref{eq10}) reduces to
\begin{equation}
\label{eq19}
a\ddot{a} - \left(\frac{\gamma - 2\omega - 1}{1 + \gamma}\right)\dot{a}^{2} = 0
\end{equation}
for a flat universe $(k = 0)$. \\
Solving Eq. (\ref{eq19}) we get the set of general solution as
\begin{equation}
\label{eq20}
a(t) = \left[\frac{2(1 + \omega)k_{3}}{(1 + \gamma)}t\right]^{\frac{(1 + \gamma)}{2(1 + \omega)}},
\end{equation}
\begin{equation}
\label{eq21}
\rho(t) = \frac{3(1 + \gamma)}{16\pi(1 + \omega)^{2}}\frac{1}{t^{2}},
\end{equation}
\begin{equation}
\label{eq22}
\Lambda(t) = \frac{3\gamma(1 + \gamma)}{2(1 + \omega)^{2}}\frac{1}{t^{2}},
\end{equation}
where $k_{3}$ is an integrating constant. In this case for physical valid solution $\gamma > 0$.
\section{Equivalence of Three Forms of $\Lambda$}

Now, let us find out the interrelations between $\alpha$, $\beta$ and $\gamma$ and hence the
equivalence of the different forms of the dynamical cosmological terms $\Lambda \sim
\left(\frac{\dot{a}}{a}\right)^2$, $\Lambda \sim \left(\frac{\ddot{a}}{a}\right)$ and
$\Lambda \sim \rho$. \\
From Eq. (\ref{eq12}), after differentiating it and the dividing by $a$, we get
\begin{equation}
\label{eq23}
t = \frac{3}{(6 - \alpha)(1 + \omega)H},
\end{equation}
where $H$ is the Hubble parameter as mentioned earlier. Therefore for specific values of $\alpha$
and $\omega$, the Eq. (\ref{eq23}) shows that $H \sim t^{-1}$. \\
Using Eq. (\ref{eq23}) in Eq. (\ref{eq13}) and also from the
definition of the cosmic matter-density parameter $\Omega(=
8\pi\rho/6H^{2})$, we get
\begin{equation}
\label{eq24}
\Omega_{m\alpha} = 1 - \frac{\alpha}{6},
\end{equation}
where $\Omega_{m\alpha}$ is the cosmic matter energy density parameter for the $\alpha$-related
dynamic $\Lambda$-model.

Using Eq. (\ref{eq23}) in (\ref{eq14}) and also from the definition of the cosmic vacuum-energy
density parameter $\Omega_{\Lambda} = \Lambda/6H^{2}$, we have
\begin{equation}
\label{eq25}
\Omega_{\Lambda \alpha} = \frac{\alpha}{6},
\end{equation}
where, in the similar fashion, $\Omega_{\Lambda \alpha}$ is the cosmic vacuum-energy density
parameter for the $\alpha$-related dynamic $\Lambda$-model.

From  Eqs. (\ref{eq24}) and (\ref{eq25}), we obtain
\begin{equation}
\label{eq26}
\Omega_{m\alpha} + \Omega_{\Lambda \alpha} = 1,
\end{equation}
which is the relation between the cosmic matter- and vacuum-energy density parameters for a
flat $(k = 0)$ universe.

Similarly from  Eqs. (\ref{eq16}) - (\ref{eq18}) we obtain
\begin{equation}
\label{eq27}
\Omega_{m\beta} = \frac{(\beta - 6)}{2(\beta + \omega \beta - 3)},
\end{equation}
\begin{equation}
\label{eq28}
\Omega_{\Lambda \beta} = \frac{\beta(1 + 2\omega)}{2(\beta + \omega \beta - 3)},
\end{equation}
where $\Omega_{m\beta}$ and $\Omega_{\Lambda \beta}$ are respectively the cosmic matter and
vacuum energy density parameters for the $\beta$-related dynamic $\Lambda$-model.

Adding (\ref{eq27}) and (\ref{eq28}), we get
\begin{equation}
\label{eq29}
\Omega_{m\beta} + \Omega_{\Lambda \beta} = 1,
\end{equation}
which is again the relation between the cosmic matter- and vacuum-energy density parameters for a
flat $(k = 0)$ universe.

In the same manner from  Eqs. (\ref{eq20}) - (\ref{eq22}) the cosmic matter and vacuum energy
density parameter $\Omega_{m\gamma}$ and $\Omega_{\Lambda \gamma}$ respectively for the model
$\Lambda \sim \rho$ also satisfy the relation
\begin{equation}
\label{eq30}
\Omega_{m\gamma} + \Omega_{\Lambda \gamma} = 1,
\end{equation}
where \\
\begin{equation}
\label{eq31}
\gamma = \frac{\Omega_{\Lambda \gamma}}{\Omega_{m\gamma}}.
\end{equation}
Thus, from the relations (\ref{eq26}), (\ref{eq29}) and (\ref{eq30}) without any loss
of generality we can set
\begin{equation}
\label{eq32}
\Omega_{m\alpha} = \Omega_{m \beta}= \Omega_{m\gamma} = \Omega_{m},
\end{equation}
\begin{equation}
\label{eq33}
\Omega_{\Lambda \alpha} = \Omega_{\Lambda \beta}= {\Omega_{\Lambda \gamma}} = \Omega_{\Lambda},
\end{equation}
where $\Omega_{m}$ and $\Omega_{\Lambda}$ are respectively the cosmic matter and vacuum energy
density parameters which in absence of any curvature satisfy the general relation
\begin{equation}
\label{eq34}
\Omega = \Omega_{m} + \Omega_{\Lambda} = 1.
\end{equation}
Now from  Eqs. (\ref{eq25}) and  (\ref{eq33}) we obtain
\begin{equation}
\label{eq35}
\alpha = 6 \Omega_{\Lambda}.
\end{equation}
Also dividing Eq. (\ref{eq27}) by (\ref{eq28}) and using Eqs. (\ref{eq32}) and  (\ref{eq33})
$\beta$ is obtained as
\begin{equation}
\label{eq36}
\beta = \frac{6 \Omega_{\Lambda}}{\Omega_{\Lambda} - (1 + 2\omega)\Omega_{m}}.
\end{equation}
Similarly from Eq. (\ref{eq31}) after using Eqs. (\ref{eq32}) and  (\ref{eq33})
\begin{equation}
\label{eq37}
\gamma = \frac{\Omega_{\Lambda}}{\Omega_{m}}.
\end{equation}
Hence from (\ref{eq34}) we get
\begin{equation}
\label{eq38} \Omega = (1 + \gamma){\Omega_{m}}=1,
\end{equation}
which is an other relation of total cosmic energy density in the case of flat universe.

All the above general relations of $ \alpha$, $ \beta$ and
$\gamma$ in terms of $\Omega_{m}$ and $\Omega_{\Lambda}$ also hold
for the particular cases of dust $(\omega=0)$ and radiation
$(\omega=1/4)$. It is interesting to note that while relation of
$\alpha$ and $\gamma$ with the cosmic matter- and vacuum-energy
density parameters are independent of $\omega$, relation of
$\beta$ with $\Omega_{m}$ and $\Omega_{\Lambda}$ depends on
$\omega$.

It can easily be shown that the particular solution of $\Lambda \sim
\left(\frac{\dot{a}}{a}\right)^2$ model for dust and radiation cases become identical with their
corresponding counterparts for the other models in term of the dependency of $a$, $\rho$ and $\Lambda$
on time $t$ when expressed in terms of $\Omega_{m}$ and $\Omega_{\Lambda} $. Therefore, these
results imply that in view of $\Omega_{m}$ and $\Omega_{\Lambda}$ there are no distinctive
features so that one can differentiate between the different forms of dynamic cosmological model
viz.$\Lambda \sim \left(\frac{\dot{a}}{a}\right)^2$, $\Lambda \sim \left(\frac{\ddot{a}}{a}\right)$
and $\Lambda \sim \rho$. Thus, starting from any one of our $\Lambda$-models, since they are
equivalent, we can arrive at the other relations.

Now from Eq. (\ref{eq35}), after using Eqs. (\ref{eq28}) and (\ref{eq33}), one can obtain
\begin{equation}
\label{eq39}
\alpha = \frac{3\beta(1 + 2\omega)}{\beta + \omega\beta - 3}.
\end{equation}
Also from Eq. (\ref{eq35}), after using Eqs. (\ref{eq37}) and  (\ref{eq38}), we get
\begin{equation}
\label{eq40} \alpha = \frac{6 \gamma}{1 + \gamma}.
\end{equation}
Thus from  Eqs. (\ref{eq39}) and  (\ref{eq40}), we find that the parameters involve in three
dynamical relations are connected by
\begin{equation}
\label{eq41} \alpha = \frac{3\beta(1 + 2\omega)}{\beta +
\omega\beta - 3} = \frac{6 \gamma}{1 + \gamma}.
\end{equation}
It is observed from above equation that the three forms $\Lambda = \alpha\left(\frac{\dot{a}}
{a}\right)^{2}$, $\Lambda = \beta\frac{\ddot{a}}{a}$ and $\Lambda = 8\pi \gamma \rho$ are
equivalent and the three parameters $\alpha$, $\beta $ and $\gamma$ are connected by the
relation (\ref{eq41}), which indicates that it is possible to find out the identical
physical features of others if any one of the phenomenological $\Lambda$ relation is known
and hence these relations are not independent of each other. Also we can see that for the
dust case Eq. (\ref{eq41}), for the relation between $\alpha$ and $\beta$, gives
\begin{equation}
\label{eq42} \alpha = \frac{3 \beta}{(\beta -3)}.
\end{equation}
In cosmology there are many ways to specify the distance between
two points, because in the expanding universe, the distance
between comoving objects are constantly changing and earth-bound
observers look back in time as they look out in distance. In the
following sections we discussed some cosmological distances for
the case $\Lambda \sim \left(\frac{\dot{a}}{a}\right)^{2}.$
\section{Neoclassical Tests (Proper Distance $d(z)$)}

A photon emitted by a source with coordinate $r=r_{1}$ and
$t=t_{1}$ and received at a time $t_{0}$ by an observer located at
$r=0$. The emitted radiation will follow null geodesics on which
$(\theta,\phi, \psi)$ are constant.

The proper distance between the source and observer is given by
\begin{equation}
\label{eq43}
 d(z)=a_{0}\int_{a}^{a_{0}}\frac{da}{a \dot{a}}\;\;\:,
\end{equation}
$$r_{1}=\int_{t_{1}}^{t_{0}}\frac{dt}{a}=\frac{a_{0}^{-1}H_{0}^{-1}A_{0}}{(1-A_{0})}
\left[ 1- (1+z)^{\frac{A_{0}-1}{A_{0}}}\right].$$ Hence
\begin{equation}
\label{eq44}
 d(z)=r_{1}a_{0}=H_{0}^{-1}\left(\frac{A_{0}}{1-A_{0}}\right)
 \left[ 1- (1+z)^{\frac{A_{0}-1}{A_{0}}}\right],
\end{equation}
where $(1 + z) = \frac{a_{0}}{a}$ = redshift and $a_{0}$ is the present scale factor of the
universe and $$A_{0} = \frac{3}{(6 - \alpha)(1 + \omega)}.$$
For small $z$ Eq. (\ref{eq44}) reduces to
 \begin{equation}
 \label{eq45}
 H_{0} d(z)= z - \frac{1}{2A_{0}}z^{2} + ...
\end{equation}
By using (\ref{eq12}) the deceleration parameter can be written as
$$q = -\frac{a\ddot{a}}{\dot{a}^{2}} = -\left(\frac{A_{0}-1}{A_{0}}\right).$$
With the use of above equation, Eq. (\ref{eq45}) can be written as
\begin{equation}
\label{eq46}
 H_{0} d(z)= z - \frac{1}{2}(1 + q)z^{2} + ...
\end{equation}
From Eq. (\ref{eq44}), it is observed that the distance $d$ is
maximum at $z = \infty$. Hence
\begin{equation}
\label{eq47}
 d(z = \infty) = H_{0}^{-1}\left(\frac{A_{0}}{1-A_{0}}\right).
\end{equation}
\section{Luminosity Distance}

Luminosity distance is the another important concept of
theoretical cosmology of a light source. The luminosity distance is a way of expanding
the amount of light received from a distant object. It is the distance that the object
appears to have, assuming the inverse square law for the reduction of light intensity
with distance holds. The luminosity distance is {\it not} the actual distance to the
object, because in the real universe the inverse square law does not hold. It is broken
both because the geometry of the universe need not be flat, and because the universe is
expanding. In other words, it is defined in such a way as generalizes the inverse-square
law of the brightness in the static Euclidean space to an expanding curved space (Waga
\cite{ref45}).

If $d_{L}$ is the luminosity distance to the object, then
\begin{equation}
\label{eq48}
d_{L} = \left(\frac{L}{4\pi l}\right)^{\frac{1}{2}},
\end{equation}
where $L$ is the total energy emitted by the source per unit time,
$l$ is the apparent luminosity of the object. Therefore one can
write
\begin{equation}
\label{eq49} d_{L} = d(1 + z).
\end{equation}
Using Eq. (\ref{eq44}), Eq. (\ref{eq49}) reduces to
\begin{equation}
\label{eq50}
 H_{0} d_{L} = (1 + z)\left(\frac{A_{0}}{1-A_{0}}\right)\Big[1 - (1 +
z)^{\frac{A_{0}-1}{A_{0}}}\Big].
\end{equation}
For small $z$, Eq. (\ref{eq50}) gives
\begin{equation}
\label{eq51}
H_{0} d_{L} = z + \frac{1}{2}(1 - q)z^{2} + ...
\end{equation}
or by using Eqs. (\ref{eq24}) and (\ref{eq32}), Eq. (\ref{eq51}) becomes
\begin{equation}
\label{eq52}
H_{0} d_{L} = z + \Big[1- (1 + \omega)\Omega_{m}\Big]z^{2} +...
\end{equation}
The luminosity distance depends on the cosmological model we have
under discussion, and hence can be used to tell us which
cosmological model describe our universe. Unfortunately, however,
the observable quantity is the radiation flux density received
from an object, and this can only be translated into a luminosity
distance if the absolute luminosity of the object is known. This
problem can however be circumvented if there are a population of
objects at different distances which are believed to have the same
luminosity; even if that luminosity is not known, it will appear
merely as an overall scaling factor.
\section{Angular Diameter Distance}

The angular diameter distance $d_{A}$ is defined as the ratio of
an object's physical transverse size to its angular size. It is
used to convert angular separations in telescope images into
proper separations at the sources.

 The angular diameter $d_{A}$ of a light source of proper distance $d$
 is given by
\begin{equation}
\label{eq53}
d_{A} = d(z)(1+z)^{-1}=d_{L}(1 + z)^{-2}.
\end{equation}
Applying Eq. (\ref{eq50})  we obtain
\begin{equation}
\label{eq54} H_{0} d_{A} =\frac{A_{0}}{1-A_{0}}\left[\frac{1 - (1
+ z)^{\frac{A_{0}-1}{A_{0}}}}{(1+z)}\right].
\end{equation}
Usually $d_{A}$ has a minimum (or maximum) for some $Z= Z_{m}$.

The angular diameter and luminosity distances have similar forms,
but have a different dependence on redshift. As with the
luminosity distance, for nearly objects the angular diameter
distance closely matches the physical distance, so that objects
appear smaller as they are put further away. However the angular
diameter distance has a much more striking behaviour for distant
objects. The luminosity distance effect dims the radiation and the
angular diameter distance effect means the light is spread over a
large angular area. This is so-called surface brightness dimming
is therefore a particularly strong function of redshift.

\section{Look Back Time}
The time in the past at which the light we now receive from a
distant object was emitted is called the look back time. How {\it
long ago} the light was emitted (the look back time)
depends on the dynamics of the universe.

The radiation travel time (or look back time) $(t - t_{0})$
for photon emitted by a source at instant $t$ and received
at $t_{0}$ is given by
\begin{equation}
\label{eq55} t - t_{0} = \int^{a_{0}}_{a} \frac{da}{\dot{a}} \; ,
\end{equation}
Equation (\ref{eq12}) can be rewritten as
\begin{equation}
\label{eq56} a = B_{0} \: t^{A_{0}}, \;\; B_{0} = constant.
\end{equation}
This follows that
\begin{equation}
\label{eq57} \frac{a_{0}}{a} = 1 + z =
\left(\frac{t_{0}}{t}\right)^{A_{0}},
\end{equation}
The above equation gives
\begin{equation}
\label{eq58}
t = t_{0}(1 + z)^{-\frac{1}{A_{0}}}.
\end{equation}
This equation can also be expressed as
\begin{equation}
\label{eq59}
H_{0} (t_{0} - t) = A_{0}\left[1 - (1 + z)^{-\frac{1}{A_{0}}}\right].
\end{equation}
For small $z$ one obtain
\begin{equation}
\label{eq60}
H_{0} (t_{0} - t) = z - \left(1+ \frac{q}{2}\right) z^{2} + ....
\end{equation}
From Eqs. (\ref{eq58}) and (\ref{eq59}), we observe that at $z \to
\infty$, $H_{0} t_{0}$ = $A_{0}$ (constant).
\section {Discussion}

In this paper we have analyzed KK type RW cosmological models by
considering three different forms of variable $\Lambda$. The
equivalence of those  forms is shown in terms of solutions thus
obtained. Since $\frac{\ddot{a}}{a} = \dot{H} + H^{2}$ then
$\Lambda \sim \frac{\ddot{a}}{a}$ models can be thought of as a
combination of two models viz. $\Lambda \sim \dot{H}$, $\Lambda
\sim H^{2}$, we observe that $\Lambda \sim \frac{\ddot{a}}{a}$ and
$\Lambda \sim \frac{\dot{a}^2}{a^2}$ models becomes identical when
$\dot{H} = 0$. Now $\dot{H} = 0$ implies $H$ is constant. In this
case we get exponential expansion and hence an inflationary
scenario. Thus the idea of inflation is inherent in
phenomenological model $\Lambda \sim \frac{\ddot{a}}{a}$. Moreover
$\Lambda \sim \frac{\dot{a}^2}{a^2}$ and $\Lambda \sim
\frac{\ddot{a}}{a}$ models can not exist as separate entity during
inflation.

We have also established the relation between $\alpha$, $\beta$ and $\gamma$, the three
parameters of the three forms of $\Lambda$ which ultimately yields $\Omega_{m} +
\Omega_{\Lambda} = 1$, the relation between cosmic matter and the vacuum energy density
parameters for flat universe. Since $\Omega_{\Lambda} = \frac{\rho_{\Lambda}}{\rho_{c}}$
and $\Omega_{m} = \frac{\rho_{m}}{\rho_{c}}$ then it is clear from Eqs. (\ref{eq35}),
(\ref{eq36}) and (\ref{eq37}) that $\beta$ and $\gamma$ are independent of $\rho_{c}$,
the critical density of the universe whereas $\alpha$ depends upon the critical density.
It can also be shown that while $\beta$ and $\gamma$ decreases with the age of universe,
$\alpha$ increases with the passage of time.

Using Eq.(\ref{eq12}) one can obtained the expression for the deceleration parameter $q$ as
$$ q = -\frac{a\ddot{a}}{\dot{a}^{2}} = \frac{(6 - \alpha)(1 + \omega)}{3} - 1. $$
Thus for an accelerating universe
$$ \alpha > \frac{3(1 + 2\omega)}{(1 + \omega)}.$$
From above equation, it is evident that for dust case, an
accelerating universe demands $\alpha > 3$. Now $\alpha$ being the
cosmic vacuum density parameter by the virtue of relation
(\ref{eq35}), we find that our model is fit for an accelerating
universe since modern accepted value of $\Omega_{\Lambda}$ is
about $0.6666$, (Turner \cite{ref63}; Frampton and Takahashi
\cite{ref64}; Perlmutter \cite{ref65}; Cardone, Troisi and
Capozziello \cite{ref66}). Thus Eq. (\ref{eq34}) implies that the
value of $\Omega_{m}$ is $0.3334$, which provides $\Omega_{m} +
\Omega_{\Lambda} = 1$ and $q= - 0.3332$, for the dust case.

It has already been shown that the three phenomenological
$\Lambda$- models presented here are equivalent, then causal
connection of the universe as indicated by the above model equally
implies that the other models of the universe are also causally
connected in the framework of higher dimensional space-time. For
the present model age of the universe is inversely proportional to
the Hubble parameter.

We also attempted to discuss the well known astrophysical
phenomena, namely the neoclassical test, the luminosity
distance-redshift, the angular diameter distance-redshift and look
back time-redshift for the model in higher dimensional space time.
It is has been observed that such models are compatible with the
results of recent observations and the cosmological term $\Lambda$
gradually reduces as the universe expands.

\section*{Acknowledgements} The authors thank the Inter-University Centre for 
Astronomy and Astrophysics (IUCAA), Pune, India, for providing warm hospitality 
and excellent facilities where this work was done. S. Otarod also thanks the 
Yasouj University for providing leave during this visit.
\newline

\end{document}